\documentclass[twocolumn]{jpsj2} 
\usepackage{multirow}

\title{
 Superconducting properties of Pr-based filled skutterudite PrRu$_{4}$As$_{12}$
}

\author{
 Takahiro \textsc{Namiki}$^{1}$\thanks{E-mail address: namiki@mmm.muroran-it.ac.jp},
 Yuji \textsc{Aoki}$^{2}$,
 Hideyuki \textsc{Sato}$^{2}$,
 Chihiro \textsc{Sekine}$^{1}$,
 Ichimin \textsc{Shirotani}$^{1}$,
 Tatsuma D. \textsc{Matsuda}$^{3}$,
 Yoshinori \textsc{Haga}$^{3}$,
 Takehiko \textsc{Yagi}$^{4}$
}

\inst{
$^{1}$ Faculty of Engineering Science, Muroran Institute of Technology, 27-1 Mizumoto, Muroran, Hokkaido 050-8585, Japan\\
$^{2}$Department of Physics, Graduate School of Science, Tokyo Metropolitan University, Minami-Ohsawa 1-1,~Hachioji, Tokyo 192-0397, Japan\\
$^{3}$Advanced Science Research Center, Japan Atomic Energy Agency, Tokai, Ibaraki 319-1195, Japan\\
$^{4}$ Institute for Solid State Physics, University of Tokyo, Kashiwa, Chiba 277-8581, Japan\\
}

\abst{
 We report on a systematic study of the superconducting characteristics and Pr crystalline-electric-field (CEF) levels of the filled skutterudite PrRu$_4$As$_{12}$ ($T_{\rm c}$ = 2.33 K).
 The temperature dependences of the upper critical field $H_{\rm c2}$ and the Ginzburg-Landau (Maki) parameter $\kappa_2$ suggest $s$-wave clean-limit superconductivity.
 The electronic specific heat coefficient $\gamma \sim 95$ mJ/K$^2$mol, which is $\sim1.5$ times larger than that of LaRu$_4$As$_{12}$, indicates $4f$-originating quasiparticle mass enhancement.
 The magnetic susceptibility $\chi(T)$ indicates that the CEF ground state is a $\Gamma_1$ singlet and a $\Gamma_4^{(1)}$ triplet first excited state lies at $\Delta_{\rm CEF}\sim 30$ \ K above.
 A systematic comparison among PrOs$_4$Sb$_{12}$, PrRu$_4$Sb$_{12}$, PrRu$_4$As$_{12}$, and La-based reference compounds suggests that the inelastic exchange scattering and aspherical charge scattering of conduction electrons from CEF-split $4f$ levels play an essential role in the quasiparticle mass enhancement and in determining the value of $T_{\rm c}$ in the Pr-based filled skutterudites.
}

\kword{Filled skutterudite, PrRu$_{4}$As$_{12}$, specific heat, susceptibility, superconductivity, crystal electric field}

\begin{document}
\maketitle

 Among heavy-fermion superconductors, the filled skutterudite PrOs$_4$Sb$_{12}$ has attracted considerable attention for its unconventional superconducting (SC) properties\cite{PRB_65_100506_2002,JPSJ_76_051006_2007_review}.
 Some of the characteristic features are broken time-reversal symmetry~\cite{PRL_91_067003_2003}, odd parity~\cite{HigemotoPRB2007}, possible structure in the SC gap~\cite{IzawaPRL2003, CustersPB2006}, multi band superconductivity~\cite{PRL_95_107004_2005}, and adjacent quadrupole ordering and its fluctuations (quadrupole excitons)~\cite{JPSJ_71_2098_2002, JPSJ_72_1002_2003, JPSJ_72_1516_2003, KuwaharaPRL2005}.
In contrast, PrRu$_4$Sb$_{12}$ is a conventional BCS-type superconductor as suggested by a coherence peak in the NQR spin-lattice relaxation rate $1/T_1$~\cite{PRB_67_180501_2003}.
Systematic understanding for such variety in the SC properties of Pr-based filled skutterudites is one of the important issues (see Table~\ref{table:SC_Parameters} for a summary of the SC parameters).

 It has been pointed out that the crystalline electric field (CEF) level scheme of Pr ions may play an essential role.
 In the $T_h$ site symmetry, the $J=4$ multiplet of Pr$^{3+}$ ions splits into four sublevels~\cite{JPSJ_70_1190_2001}, namely, a singlet $\Gamma_1$, a non-Kramers nonmagnetic doublet $\Gamma_{23}$, and two triplets $\Gamma_{4}^{(1)}$ and $\Gamma_{4}^{(2)}$.
 PrOs$_4$Sb$_{12}$ has $\Gamma_1$ ground state with a $\Gamma_{4}^{(2)}$ first excited state with the energy separation $\Delta_{\rm CEF}=8$ K~\cite{GoremychkinINSPOSPRL2004,KuwaharaPRL2005}.
 In contrast, PrRu$_4$Sb$_{12}$ has $\Gamma_1- \Gamma_{4}^{(1)}$ levels with $\Delta_{\rm CEF}=65$ K\cite{JPSJ_69_868_2000,PB_359-361_983_2005,PB_359-361_983_2005_Comment}.
 The differences in the type of the first excited state and $\Delta_{\rm CEF}$ may be a key reason for the different SC properties.

 In this paper, we report the specific heat and magnetic susceptibility measurements of another Pr-based filled skutterudite PrRu$_4$As$_{12}$~\cite{PRB_56_7866_1997} to study the SC properties and the CEF level scheme.
The results indicate that the superconductivity is of the conventional BCS type with a moderate quasiparticle mass enhancement.
 A systematic comparison among these Pr-based compounds and their La-based references  reveals that inelastic scatterings of conduction electrons from CEF-split $4f$ levels play an essential role for the  SC properties and quasiparticle mass enhancement in the Pr-based filled skutterudites.

 Polycrystalline PrRu$_4$As$_{12}$ samples are synthesized using stoichiometric amounts of 3N(99.9\% pure)-Pr, 4N-Ru, and 5N-As powders by the high-pressure and high-temperature method ($\sim4$ GPa at 900 $^\circ\mathrm{C}$)~\cite{RHPST_6_109_1997}. 
 Powder x-ray diffraction indicates the inclusion of a small amount of the RuAs$_{2}$ phase; the maximum peak height of RuAs$_{2}$ is approximately 10 \% of that of PrRu$_4$As$_{12}$.
The structural parameters of PrRu$_4$As$_{12}$ are determined by single-crystal x-ray diffraction using an imaging plate detector with Mo K$_{\alpha}$ radiation.
 A small single crystal with approximate dimensions of 0.10 $\times$ 0.07 $\times$ 0.05 mm is selected from the ingot.
 The structural parameters are successfully refined based on the 179 independent reflections with the agreement factor $\sum{ \Bigl| \left| F_{\rm o} \right| - \left| F_{\rm c} \right| \Bigr| }/ \sum {\left|F_{\rm o} \right|  }$ = 0.023, where $F_{\rm o}$ and $F_{\rm c}$ are the experimental and calculated structure factors, respectively.
 Fractional coordinates and Debye-Waller factors ($U_{\rm eq}$) are listed in Table~\ref{table:PrRu4As12_Posistion}.
\begin{table}
 \caption{
 Fractional coordinates and Debye-Waller factors $U_{\rm eq}$ of PrRu$_4$As$_{12}$ determined from the single-crystal x-ray diffraction data.
 }
 \label{table:PrRu4As12_Posistion}
 \begin{center}
  \begin{tabular}{cccccc}
   \hline
   atom & site  & $x$ & $y$        & $z$       & $U_{\rm eq}({\rm \AA}^2)$ \\
   \hline
   Pr   & 2$a$  &  0  &  0         &  0        & 1.082(4)                  \\
   Ru   & 8$c$  & 1/4 & 1/4        & 1/4       & 0.294(3)                  \\
   As   & 24$g$ &  0  & 0.14954(3) & 0.35019(3)& 0.387(5) \\
   \hline
  \end{tabular}
 \end{center}
\end{table}
 The obtained lattice constant of 8.491(4) \AA\ agrees with a reported value~\cite{JSSC_32_357_1980}.
$U_{\rm eq}$(Pr) is $\sim 3$ times larger than $U_{\rm eq}$(Ru) and $U_{\rm eq}$(As).
 Note that $U_{\rm eq}$(Pr) in PrRu$_4$As$_{12}$ is larger than that in PrRu$_{4}$P$_{12}$~\cite{PB_322_408_2002}, indicating larger thermal vibrations of Pr ions in As$_{12}$ cages, whose size is larger than that of P$_{12}$ cages.

 The specific heat $C(H,T)$ is measured by a quasiadiabatic heat pulse method~\cite{JPSJ_71_2098_2002} using a dilution refrigerator equipped with an 8 T SC magnet.
 The temperature increment caused by each heat pulse is controlled to $\sim$2 \% for the usual measurement and to $\sim$0.5 \% in limited temperature ranges where the SC transition occurs.
 The magnetic susceptibility $\chi$ is measured in $1.8<T<300$ K using an SC quantum interference device (SQUID) magnetometer (Quantum Design Inc.).

\begin{figure}
 \begin{center}
 \includegraphics[width=0.7\linewidth]{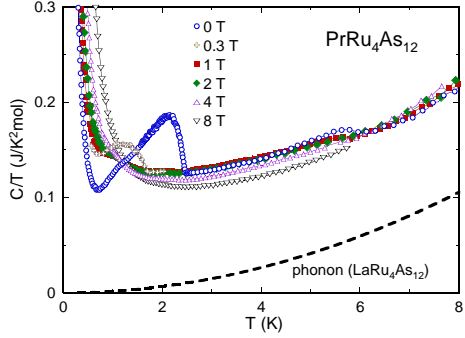}
 \end{center}
\caption{
 (Color online)
 Specific heat divided by temperature ($C/T$) of PrRu$_4$As$_{12}$ measured in several magnetic fields.
 The broken curve represents the phonon part $C_{\rm ph}/T$ determined for LaRu$_4$As$_{12}$~\cite{PRB_56_7866_1997}.
}
\label{fig:PrRu4As12_Poli_CbT-vs-T_All}
\end{figure}

\begin{figure}[htbp]
 \begin{center}
 \includegraphics[width=0.7\linewidth]{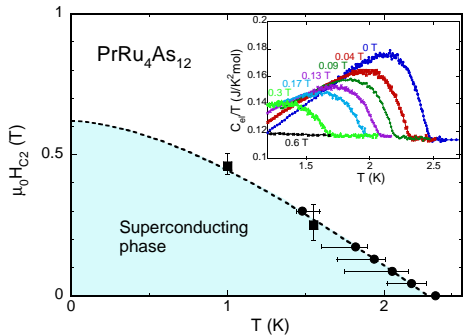}
 \end{center}
\caption{
 (Color online)
Upper critical field $\mu_0H_{\rm c2}$ vs $T$ determined by $T$-sweep and $H$-sweep measurements of $C$.
 The error bars denote the 10\% -- 90\% width of the specific heat jump.
 The broken line shows the best-fit WHH curve (see text).
 The inset shows $C_{el}/T$ vs $T$ near $T_{\rm c}$.
}
\label{fig:PrRu4As12_Poli_H-vs-T_PhaseDiagram_Tc}
\end{figure}

 The temperature dependence of $C/T$ measured in selected magnetic fields is shown in Fig.~\ref{fig:PrRu4As12_Poli_CbT-vs-T_All}.
 In zero field, a clear jump appears at the SC transition temperature $T_{\rm c}$ = 2.33 K, which is very close to a resistively determined value of 2.4 K\cite{PRB_56_7866_1997}.
 With increasing field, $T_{\rm c}$ shifts to lower temperatures and the size of the jump becomes smaller (for details, see the inset of Fig.~\ref{fig:PrRu4As12_Poli_H-vs-T_PhaseDiagram_Tc}).

 In the measured temperature range, $C$ consists of three terms, namely, the electronic contribution ($C_{\rm el}$), phonon contribution ($C_{\rm ph}$), and nuclear Schottky contribution ($C_{\rm n}$).
 In zero field, $C_{\rm n}= A(H)/T^2$ appears as a low-$T$ upturn below 0.7 K.
 This term gradually develops with increasing field.
 The phonon term $C_{\rm ph}=\beta T^3$ with $\beta = 1.53$ mJ/K$^4$mol determined for LaRu$_4$As$_{12}$~\cite{PRB_56_7866_1997} is indicated by the broken curve in Fig.~\ref{fig:PrRu4As12_Poli_CbT-vs-T_All}.
 By the fitting using $C(T,H) = C_{\rm el}(T,H) + A(H)/T^2 + \beta T^3$, $C_{\rm el}(T,H)$ is extracted (see the inset of Fig.~\ref{fig:PrRu4As12_Poli_H-vs-T_PhaseDiagram_Tc}).

 Note that there appears a slight jump at $T_{\rm A} \sim 6$ K in zero field.
This anomaly is probably due to a magnetic ordering in a secondary phase included in the present sample since no corresponding anomalies appear in recent $^{75}$As-NQR measurements\cite{submit_Shimizu}.
 This phase has not been identified by the x-ray diffraction measurements.
 To synthesize a sample free from the secondary phase, sample preparation is conducted several times under different conditions and the best sample is used for the present measurements.
 Probably due to the contamination of the secondary phase, $C/T$ decreases with increasing field in the normal state, making it difficult to estimate the electronic specific heat coefficient ($\gamma$).
 Nevertheless, from the $C$ data at 8 T, where the contribution of the secondary phase is largely suppressed, $\gamma \sim 95$ mJ/K$^2$mol is estimated.
 The specific heat jump $\Delta C/\gamma T_{\rm c} \sim 0.83$, smaller than 1.43 for a weak-coupling BCS superconductor, and apparent $\sim T^2$ dependence in $C(T)$ below $T_{\rm c}$ may suggest SC gap anisotropy or multi band superconductivity.  
 However, a clear coherence peak observed in As-NQR $1/T_1$ indicates that such nonuniformity would not be very large~\cite{submit_Shimizu}.

 From the $C_{\rm el}(T, H)$ data, the upper critical field $\mu_0H_{\rm c2}(T)$ is determined, as shown in the $H$-vs-$T$ phase diagram of Fig.~\ref{fig:PrRu4As12_Poli_H-vs-T_PhaseDiagram_Tc}.
 It appears that the $\mu_0H_{\rm c2}(T)$ curve is well reproduced by the Werthamer-Hefland-Hohemberg (WHH) formula for the clean limit~\cite{PR_147_288_1966,PR_147_295_1966} with the initial slope $[-d(\mu_0H_{\rm c2})/dT]_{T_{\rm c}} = 0.365$ T/K.
 This fact indicates that $\mu_0H_{\rm c2}$ is mainly determined by the orbital effect and the spin Pauli paramagnetic effect does not play a dominant role; the Pauli limiting field is $\mu_0H_{\rm P}=3.9$ T $\ll \mu_0H_{\rm c2}$.
 The $T=0$ value of $\mu_0H_{\rm c2}$ is estimated to be $\sim0.62$ T by the WHH fitting.
\begin{table*}[tb]
 \caption{
 Superconducting parameters of Pr-based filled skutterudites and their isostructural La-based reference compounds.
 }
 \label{table:SC_Parameters}
  \begin{tabular}{ccccccccc}
  \hline
\scriptsize  Material & \scriptsize $T_{\rm c}$ & \scriptsize $\mu_0H_{\rm c2}$ & \scriptsize {[$-\frac{dH_{\rm c2}}{dT}$]$_{T_c}$} & \scriptsize $\Delta_{\rm CEF}$ & \scriptsize First excited state & \scriptsize $\gamma$ & \scriptsize $m_c^*$/$m_0$ for $\gamma$-branch & Reference\\
       & \scriptsize  (K) & \scriptsize  (T) & \scriptsize (T/K) & \scriptsize (K) & \scriptsize                  & \scriptsize (mJ/K$^2$mol) & \scriptsize     & \scriptsize   \\
  \hline
  PrRu$_4$As$_{12}$ & \scriptsize 2.33 & \scriptsize 0.62  & \scriptsize 0.37  & \scriptsize $\sim 30$  & \scriptsize $\Gamma_4^{(1)}$ & \scriptsize 95            & \scriptsize     & \scriptsize This work \\
  LaRu$_4$As$_{12}$ & \scriptsize 10.3 & \scriptsize 0.72 & \scriptsize 0.08  & \scriptsize     & \scriptsize                  & \scriptsize 73            & \scriptsize     & \scriptsize \citen{PRB_56_7866_1997} \\
  \hline
  PrRu$_4$Sb$_{12}$ & \scriptsize 1.03 & \scriptsize 0.2\ & \scriptsize 0.24  & \scriptsize 65  & \scriptsize $\Gamma_4^{(1)}$ & \scriptsize 59            & \scriptsize 1.6 & \scriptsize \citen{JPSJ_69_868_2000,PB_312-313_832_2002,PB_359-361_983_2005} \\
  LaRu$_4$Sb$_{12}$ & \scriptsize 3.58 & \scriptsize 0.28 & \scriptsize 0.12  & \scriptsize     & \scriptsize                  & \scriptsize 37            & \scriptsize 1.4 & \scriptsize \citen{JPSJ_69_868_2000,PB_359-361_983_2005} \\
  \hline
  PrOs$_4$Sb$_{12}$ & \scriptsize 1.81 & \scriptsize 2.3  & \scriptsize 1.9   & \scriptsize  8  & \scriptsize $\Gamma_4^{(2)}$ & \scriptsize $\sim$ 500    & \scriptsize 7.6 & \scriptsize \citen{PRB_65_100506_2002,KuwaharaPRL2005,PRB_66_220504_2002} \\
  LaOs$_4$Sb$_{12}$ & \scriptsize 0.74 & \scriptsize 0.04 & \scriptsize 0.095 & \scriptsize     & \scriptsize                  & \scriptsize 55            & \scriptsize 2.8 & \scriptsize \citen{JPCM_13_4495_2001,PRB_66_220504_2002} \\
  \hline
 \end{tabular}
\end{table*}

 There appears a slight upward deviation in the $\mu_0H_{\rm c2}(T)$ curve below 0.05 T, leading to an enhancement in $T_{\rm c}$ by 2\% in zero field compared with the best-fit WHH curve for $\mu_0H>0.05$ T.
 A similar behavior is observed in the case of PrOs$_4$Sb$_{12}$\cite{PRB_70_064516_2004} and YNi$_{2}$B$_{2}$C\cite{PRL_80_1730_1998}, indicating that the upward deviation in $T_{\rm c}$ may come from the multi band effect.

\begin{figure}[htbp]
 \begin{center}
 \includegraphics[width=0.7\linewidth]{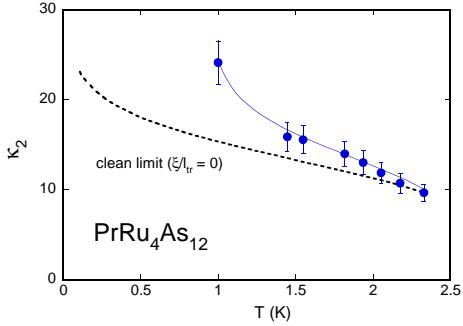}
 \end{center}
\caption{
 Temperature dependence of Ginzburg-Landau parameter (Maki parameter) $\kappa_2$ for PrRu$_4$As$_{12}$.
 The broken curve indicates the theoretical curve for the clean limit.
 The solid curve is a guide to the eye.
}
\label{fig:PrRu4As12_Poli_kappa2-vs-T_All}
\end{figure}

 By combining the $\mu_0H_{\rm c2}$ and $\Delta C/T_{\rm c}$ data, the Ginzburg-Landau parameter (or Maki parameter)~\cite{PR_139_A868_1965} $\kappa_2$ is determined using the thermodynamical relation:
\begin{equation}
\label{eq:Maki_kappa2}
 \Delta C/T_{\rm c} = \left[\frac{d\left(\mu_0H_{{\rm c2}}\right)}{dT_{\rm c}} \right] \times \frac{1}{4\pi\left(2\kappa^2_2 - 1\right)\beta_A},
\end{equation}
 where $\beta_A$ is 1.16 for a triangular vortex lattice~\cite{SPJETP_5_1174_1957}.
 The result is shown in Fig.~\ref{fig:PrRu4As12_Poli_kappa2-vs-T_All}.
The value of $\kappa_2$ is $\sim$ 10 at $T=T_{\rm c}$ (in zero field) and increases with decreasing temperature~\cite{PrRu4As12_Comment01}.
The behavior of $d \kappa_2/dT<0$ is again consistent with the negligible Pauli paramagnetic depairing effect mentioned above; for superconductors with a strong Pauli paramagnetic effect, $d \kappa_2/dT>0$ appears (e.g., CeCoIn$_{5}$; see Ref.~\citen{JPSJ_70_2248_2001}).
 The temperature dependence of $\kappa_2$ is affected by impurity scatterings~\cite{PRB_68_184503_2003}.
 In the clean limit, the theoretical parameter $\kappa_2(T)$ diverges as $T \rightarrow$ 0 with $\kappa_2 \propto \sqrt{\ln(T_{\rm c}/T)}$~\cite{PR_153_584_1967,PR_139_A868_1965} as shown in Fig.~\ref{fig:PrRu4As12_Poli_kappa2-vs-T_All}.
 However, in the dirty limit, the temperature dependence of $\kappa_2$ is largely suppressed ($\kappa_2(0)/\kappa_2(T_{\rm c}) \sim 1.2$).
 The $\kappa_2(T)$ data in Fig.~\ref{fig:PrRu4As12_Poli_kappa2-vs-T_All} indicates that the present sample is rather close to the clean limit.

\begin{figure}[htbp]
 \begin{center}
 \includegraphics[width=0.7\linewidth]{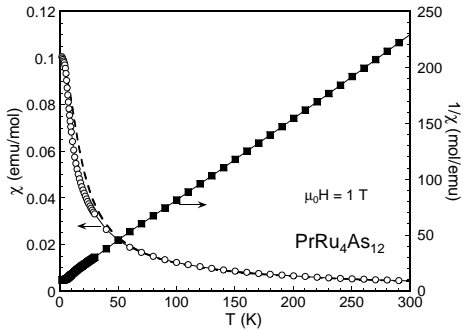}
 \end{center}
\caption{
 Temperature dependences of susceptibility $\chi$ (open circle) and inverse susceptibility $\chi^{-1}$ (open square).
 The dashed line shows the best-fit CEF model calculation (see text).
}
\label{fig:PrRu4As12_Poli_MbH-vs-T_1T}
\end{figure}

 The magnetic susceptibility $\chi(T)$ is shown in Fig.~\ref{fig:PrRu4As12_Poli_MbH-vs-T_1T}.
 At high temperatures, $\chi(T)$ can be well described by a Curie-Weiss law $\chi = N_A \mu_{eff}^{2}/3k_B(T-\Theta_{CW})$, where the effective magnetic moment $\mu_{eff} = 3.30  \mu_{\rm B}$/f.u. and the Curie-Weiss temperature $\Theta_{CW}$ = --11 K.
 The obtained $\mu_{eff}$ value is close to the Pr$^{3+}$ free-ion value (3.58 $\mu_{\rm B}$/Pr), indicating a well-localized character of Pr $4f$ electrons.
 Below $\sim$10 K, $\chi(T)$ shows a deviation from the Curie-Weiss law, exhibiting a saturation tendency.
 This behavior suggests that the low-lying CEF level scheme of Pr$^{3+}$ ions has a $\Gamma_1$ singlet ground state and a $\Gamma_4^{(1)}$ triplet first excited state.
 Note that in the case of a $\Gamma_4^{(2)}$ first excited state, as realized in PrOs$_4$Sb$_{12}$, a maximum should appear in $\chi(T)$~\cite{PRB_65_100506_2002,JPSJ_71_2098_2002}.
 This interpretation agrees with the electronic entropy $S_{\rm el}(T,H)$ (not shown) calculated from the $C_{\rm el}(T,H)$ data.
$S_{\rm el}$ at 8 K is less than 20 \% of $R\ln 2$, ruling out the possibility of any degenerate ground states.
 In the measured temperature range, $dS_{\rm el}/dH$ is always negative.
 In PrOs$_4$Sb$_{12}$, $dS_{\rm el}/dH>0$ is observed below 3 K~\cite{JPSJ_71_2098_2002}.
 This behavior is caused by the Zeeman splitting of the low-lying triplet first excited state at $\Delta_{\rm CEF}=8 $ K. This fact indicates that $\Delta_{\rm CEF} \gg 8$ K for PrRu$_4$As$_{12}$.
 For a rough estimation of the low-lying CEF level scheme, a least-squares fit has been made based on the $O_h$ Lea-Leask-Wolf scheme~\cite{JPCS_23_1381_1962} (omitting the $O_6^2-O_6^6$ term characteristic of the $T_h$ site symmetry in the CEF Hamiltonian).
 The best overall fit of the $\chi(T)$ data (shown by the dashed line in Fig.~\ref{fig:PrRu4As12_Poli_MbH-vs-T_1T}) corresponds to a $\Gamma_1$ ground state with a $\Gamma_4$ ($\Gamma_4^{(1)}$ in $T_h$) triplet excited state at $\Delta_{\rm CEF} \sim 30$ K~\cite{comment_PrRu4As12_CEF}.  
 For an accurate determination of the overall CEF level scheme, inelastic neutron scattering measurements would be necessary.

\begin{figure}[htbp]
 \begin{center}
 \includegraphics[width=0.7\linewidth]{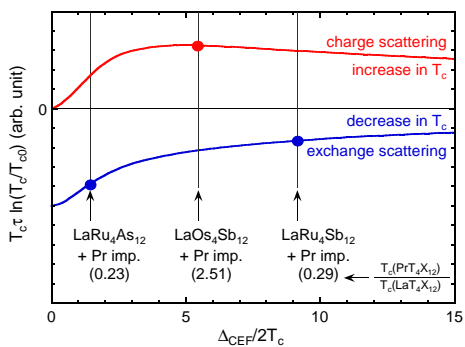}
 \end{center}
\caption{
 (Color online)
 Two curves showing calculated changes in $T_{\rm c}$ vs $\Delta_{\rm CEF}/2T_{\rm c}$ by doping impurity ions with singlet-singlet CEF levels (see Ref.~\citen{ZP_230_155_1970} for the parameter definition).
 For a rough estimation of the Pr doping effect on $T_{\rm c}$, $\Delta_{\rm CEF}/2T_{\rm c}$ for LaT$_4$X$_{12}$ is indicated.
 The expected dominant scattering process is indicated by solid circles.
}
\label{fig:Fulde-Calc}
\end{figure}

 Only for LaOs$_4$Sb$_{12}$, among the three systems shown in table~\ref{table:SC_Parameters}, an enhancement in $T_{\rm c}$ is caused by the replacement of La by Pr ions.
This behavior may be accounted for by considering the predominant type of conduction-electron scatterings from Pr $4f$ electrons.
 Fulde {\it et al.}~\cite{ZP_230_155_1970} have theoretically demonstrated that inelastic aspherical charge scattering (ACS) associated with $4f$ quadrupole moments leads to an increase in $T_{\rm c}$, while exchange scattering (ES) decreases $T_{\rm c}$.
 For a qualitative discussion, a comparison with the theoretical calculation is made in Fig.~\ref{fig:Fulde-Calc}.
 For LaOs$_4$Sb$_{12}$, since the first excited state of the replaced Pr ions is $\Gamma_4^{(2)}$ ($\sim \Gamma_5$ in $O_h$ having large off-diagonal quadrupole moments with $\Gamma_1$), the ACS may dominate over the ES.
 Furthermore, $\Delta_{\rm CEF}/2T_{\rm c}($LaOs$_4$Sb$_{12}$$) \simeq 5$ happens to be the best condition for the enhancement in $T_{\rm c}$.
 On the other hand, in LaRu$_4$As$_{12}$ and LaRu$_4$Sb$_{12}$, the first excited state of the 
replaced Pr ions is $\Gamma_4^{(1)}$ ($\sim \Gamma_4$ in $O_h$ having large 
off-diagonal dipole moments with $\Gamma_1$) and the resulting predominant ES 
will cause a strong pair-breaking effect, thereby decreasing $T_{\rm c}$.

\begin{figure}[htbp]
 \begin{center}
 \includegraphics[width=0.7\linewidth]{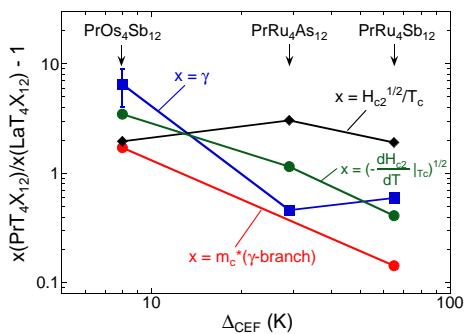}
 \end{center}
\caption{
 (Color online)
Mass enhancement factor $m^*/m_0-1$, estimated from several physical quantities, is plotted as a function of $\Delta_{\rm CEF}$.
}
\label{fig:MassEnhancement-vs-CEF}
\end{figure}

 The quasiparticle mass enhancement is also expected to be caused by the interactions of conduction electrons with low-lying CEF levels.
 Flude \textit{et al.} have proposed a model for the mass enhancement due to virtual CEF excitations~\cite{PRB_27_4085_1983}:
\begin{equation}
 \frac{m^*}{m_0} -1 = (g_J-1)^2J_{sf}^2 N(0)\frac{2\left|\langle i\left|J\right|j\rangle\right|^2}{\Delta_{\rm CEF}},
 \label{eq:mass_enhancement}
\end{equation}
 where $g_J$ is the Land\'{e} factor, $J_{sf}$ is the exchange integral coupling the conduction electrons to the $f$ electrons, $N(0)$ is the bare conduction electron density of states at the Fermi level, and $\langle i\left|J\right|j\rangle$ is the magnetic dipole matrix element calculated using the derived CEF parameters.
 Figure~\ref{fig:MassEnhancement-vs-CEF} shows several physical quantities that provide measures of the mass enhancement as a function of $\Delta_{\rm CEF}$.
 There appears a rough trend of $m^*/m-1 \propto 1/\Delta_{\rm CEF}$ in Fig. 6.
 However, as discussed above for the effect on $T_{\rm c}$, not only the ES process but also the ACS process should contribute to the mass enhancement with a different prefactor of $1/\Delta_{\rm CEF}$.
 In PrOs$_4$Sb$_{12}$, the importance of quadrupolar degrees of freedom is inferred from a strong correlation between quadrupolar excitons and superconductivity~\cite{KuwaharaPRL2005}.
 Further experimental and theoretical studies on the role of quadrupoles in Pr-based filled skutterudites will be needed.

 We thank M. Ichioka, Y. Kohori, H. Kusunose, N. K. Sato, K. Takeda, and Y. Yanase for fruitful discussions.
 This work was partially supported by a Grant-in-Aid for Scientific Research Priority Area ``Skutterudite'' (Nos. 15072101, 15072206, 18027002) of MEXT, Japan  Scientific Research (c) (No. 17540309) of MEXT, Japan.
 



\begin{thebibliography}{10}
\bibitem{PrRu4As12_Comment01}
 Note that the inclusion of the secondary phase does not significantly affect the overall $T$-dependent behavior of $\kappa_2$, although the resulting possible underestimation of $\Delta C$ may have led to the overestimation of $\kappa_2$ to some extent.
\bibitem{comment_PrRu4As12_CEF}
 The fitting suggests that the other excited states lie probably above $\sim 50$ K in energy. Note that $\chi(T)$ is not reliable for the determination of high-energy CEF excitations.
\end{thebibliography}


\end{document}